\documentclass[twocolumn,showpacs,superscriptaddress,amsmath,amssymb,aps,prb]{revtex4-1}
\usepackage{bm}
\usepackage{graphicx}
\usepackage{dcolumn}
\usepackage{amsmath,txfonts}
\usepackage{epstopdf}
\usepackage[mathlines]{lineno}
\usepackage{color}
\usepackage[colorlinks=true,linkcolor=blue,citecolor=blue]{hyperref}
\usepackage[normalem]{ulem}

\begin{document}

\title{Higher-order exceptional point in a cavity magnonics system}

\author{Guo-Qiang Zhang}
\affiliation{Quantum Physics and Quantum Information Division, Beijing Computational Science Research Center, Beijing 100193, China}
\affiliation{Interdisciplinary Center of Quantum Information and Zhejiang Province Key Laboratory of Quantum Technology and Device, Department of Physics and State Key Laboratory of Modern Optical Instrumentation, Zhejiang University, Hangzhou 310027, China}

\author{J. Q. You}
\email{Corresponding author. jqyou@zju.edu.cn}
\affiliation{Interdisciplinary Center of Quantum Information and Zhejiang Province Key Laboratory of Quantum Technology and Device, Department of Physics and State Key Laboratory of Modern Optical Instrumentation, Zhejiang University, Hangzhou 310027, China}

\begin{abstract}
We propose to realize the pseudo-Hermiticity in a cavity magnonics system consisting of the Kittel modes in two small yttrium-iron-garnet spheres coupled to a microwave cavity mode. The effective gain of the cavity can be achieved using the coherent perfect absorption of the two input fields fed into the cavity. With certain constraints of the parameters, the Hamiltonian of the system has the pseudo-Hermiticity and its eigenvalues can be either all real or one real and other two constituting a complex-conjugate pair. By varying the coupling strengths between the two Kittel modes and the cavity mode, we find the existence of the third-order exceptional point in the parameter space, in addition to the usual second-order exceptional point existing in the system with parity-time symmetry. Also, we show that these exceptional points can be demonstrated by measuring the output spectrum of the cavity.
\end{abstract}

\date{\today}

\maketitle

\section{Introduction}

By harnessing the advantages of different components, the hybrid quantum systems have potential applications in quantum information.~\cite{Xiang13,Kurizki15} Among various hybrid systems, the cavity magnonics system has received increasing interest in recent years;~\cite{SoykalPRL10,SoykalPRB10,Huebl13,Tabuchi14,Zhang14,Zhang15-1,Rameshti15,
Cao15,Liu16,Sharma17,Sharma18,Osada18,Grigoryan18,Liu18} here \emph{magnonics} is related to an emergent branch of magnetism, with the main aim to investigate the behavior of spin waves in a confined or nanostructured system.~\cite{Kruglyak10} In such a hybrid system, magnons in, e.g., a small yttrium iron garnet (YIG) sample are coupled to microwave photons in a cavity. Originating from the high spin density and the strong spin-spin exchange interactions, the Kittel mode in the YIG sample can possess both a long coherence time and a low damping rate,~\cite{Zhang15-1,Cherepanov93} making the cavity magnonics system easy to reach the strong-coupling regime~\cite{Huebl13,Tabuchi14,Zhang14} and even possible to reach the ultrastrong-coupling regime.~\cite{Bourhill16,Kostylev16} Moreover, owing to the merits of high tunability and good coherence, the cavity magnonics system has become a promising platform to implement various novel phenomena, such as the magnon gradient memory,~\cite{Zhang15-2} bistability of cavity-magnon polaritons,~\cite{Wang16,Wang18} cavity spintronics~\cite{Bai15,Bai17} and cooperative polariton dynamics.~\cite{Yao17} In addition, it was experimentally shown that the magnons in the small YIG sample can couple to the optical photons,~\cite{Haigh15,Osada16,Zhang16-2,Haigh16} phonons,~\cite{Zhang16-1} and superconducting qubit.~\cite{Tabuchi15,Quirion17} This makes it promising to produce the magnon-photon-phonon entanglement in cavity magnomechanics.~\cite{Agarwal}

As stated in quantum mechanics, the Hamiltonian of a closed quantum system must be Hermitian to have a real energy spectrum. However, any realistic quantum systems are actually open systems. Under certain conditions, they may be effectively modeled by the non-Hermitian Hamiltonians. In Refs.~\onlinecite{Mostafazadeh02-1,Mostafazadeh02-2,Mostafazadeh02-3}, Mostafazadeh proposed the pseudo-Hermiticity for the non-Hermitian Hamiltonian of the system: If a Hamiltonian $H$ with a discrete spectrum satisfies $H^{\dag}=U H U^{-1}$, where $\dag$ denotes the Hermitian adjoint and $U$ is a linear Hermitian operator, the Hamiltonian $H$ is pseudo-Hermitian and its eigenvalues are either real or complex conjugate pairs. The pseudo-Hermiticity is an interesting topic in non-Hermitian physics, which can give rise to rich exotic phenomena in different subjects of physics (e.g., quantum chaos and quantum phase transitions,~\cite{Deguchi09-1,Deguchi09-2} Dirac particles in gravitational fields,~\cite{Gorbatenko10} Maxwell's equations,~\cite{Zhu11} anisotropic \emph{XY} model,~\cite{Zhang13} and dynamical invariants~\cite{Simeonov16}).

Obviously, the Hermiticity of the Hamiltonian, $H^{\dag}=H$, is a special case of the pseudo-Hermiticity, with $U$ being a unit operator. Also, the $\mathcal{PT}$-symmetric Hamiltonian is another subset of the pseudo-Hermitian Hamiltonian,~\cite{Mostafazadeh02-3} where the Hamiltonian $H$ satisfies~\cite{Konotop16} $[H,\mathcal{PT}]=0$, with $\mathcal{P}$ and $\mathcal{T}$ being the parity and time operators, respectively. Hereafter, the pseudo-Hermiticity mentioned below excludes both the Hermiticity and the $\mathcal{PT}$ symmetry. When varying one of the system's parameters near the critical point (i.e., the exceptional point), the system undergoes a quantum phase transition from the $\mathcal{PT}$-symmetric phase to the $\mathcal{PT}$-symmetry-breaking phase (with real and complex eigenvalues, respectively) in the parameter space.~\cite{Konotop16} This exceptional point is also called the second-order exceptional point ($\text{EP}_{2}$)
and has been studied in various non-Hermitian systems, including the optomechanical systems,~\cite{Xu16,Lu15} coupled waveguides,~\cite{Doppler16} coupled optical microresonators,~\cite{Chang14} cavity magnonics systems,~\cite{Zhang17,Harder17,Gao17,Wang181} and superconducting circuit-QED systems.~\cite{Quijandria18} Besides $\text{EP}_{2}$, high-order exceptional points may occur in non-Hermitian systems. Specifically, an $n$th-order exceptional point ($\text{EP}_{n}$) corresponds to the coalescence of $n$ eigenvalues in a non-Hermitian linear system.~\cite{Heiss12} Higher-order exceptional points are more complicated but can exhibit richer physical phenomena.~\cite{Lin16,Graefe08,Ryu12,Heiss15,Schnabel17,Jing17,Wu18} For instance, a higher-order exceptional point has much richer topological characteristics in coupled acoustic resonators~\cite{Ding16} and can further enhance the sensitivity of the sensors in photonic molecules.~\cite{Hodaei17} To the best of our knowledge, there is no study on both the pseudo-Hermiticity without the $\mathcal{PT}$ symmetry and the related higher-order exceptional point in a cavity magnonics system.

In this work, we investigate the high-order exceptional point in a cavity magnonics system by designing an effective pseudo-Hermitian Hamiltonian without the $\mathcal{PT}$ symmetry. In our proposal, the hybrid system is composed of two small YIG spheres placed in a microwave cavity, where the Kittel mode in each YIG sphere is strongly coupled to the cavity mode. In order to realize the pseudo-Hermiticity of the Hamiltonian, a gain of the cavity is needed, which can be effectively achieved using the coherent perfect absorption (CPA) of the two input fields fed into the cavity via two ports.~\cite{Zhang17,Sun14}
In addition to the usual $\text{EP}_{2}$, we find the third-order exceptional point ($\text{EP}_{3}$) in the parameter space. Moreover, we show that the $\text{EP}_{3}$ can be observed via measuring the total output spectrum of the cavity, where the CPA frequencies are found to be coincident with the real energy spectrum of the hybrid system.

Our work brings the study of cavity magnonics systems to the interesting pseudo-Hermitian physics. In previous works,~\cite{Zhang17,Sun14,Heiss12,Xu16,Lu15,Doppler16,Chang14,Harder17,Gao17,Wang181,Quijandria18,Lin16,Graefe08,Ryu12,Heiss15,Schnabel17,Jing17,Wu18,Ding16,Hodaei17} exceptional points were realized in either the $\mathcal{PT}$-symmetric system or the non-Hermitian system without the pseudo-Hermiticity. Our work provides an initial study to the high-order exceptional point in a cavity magnonics system owning the pseudo-Hermitian Hamiltonian without the $\mathcal{PT}$ symmetry. In contrast to Ref.~\onlinecite{Zhang17}, we design a more sophisticated system and show that the CPA can also occur in the absence of the $\mathcal{PT}$ symmetry. Also, our proposed hybrid system may be harnessed to explore exotic phenomena of the high-order exceptional point (e.g., the topological properties~\cite{Ding16} and the perturbation amplification~\cite{Hodaei17}) in the future.

\section{The Model}\label{model}

\begin{figure}
\includegraphics[width=0.44\textwidth]{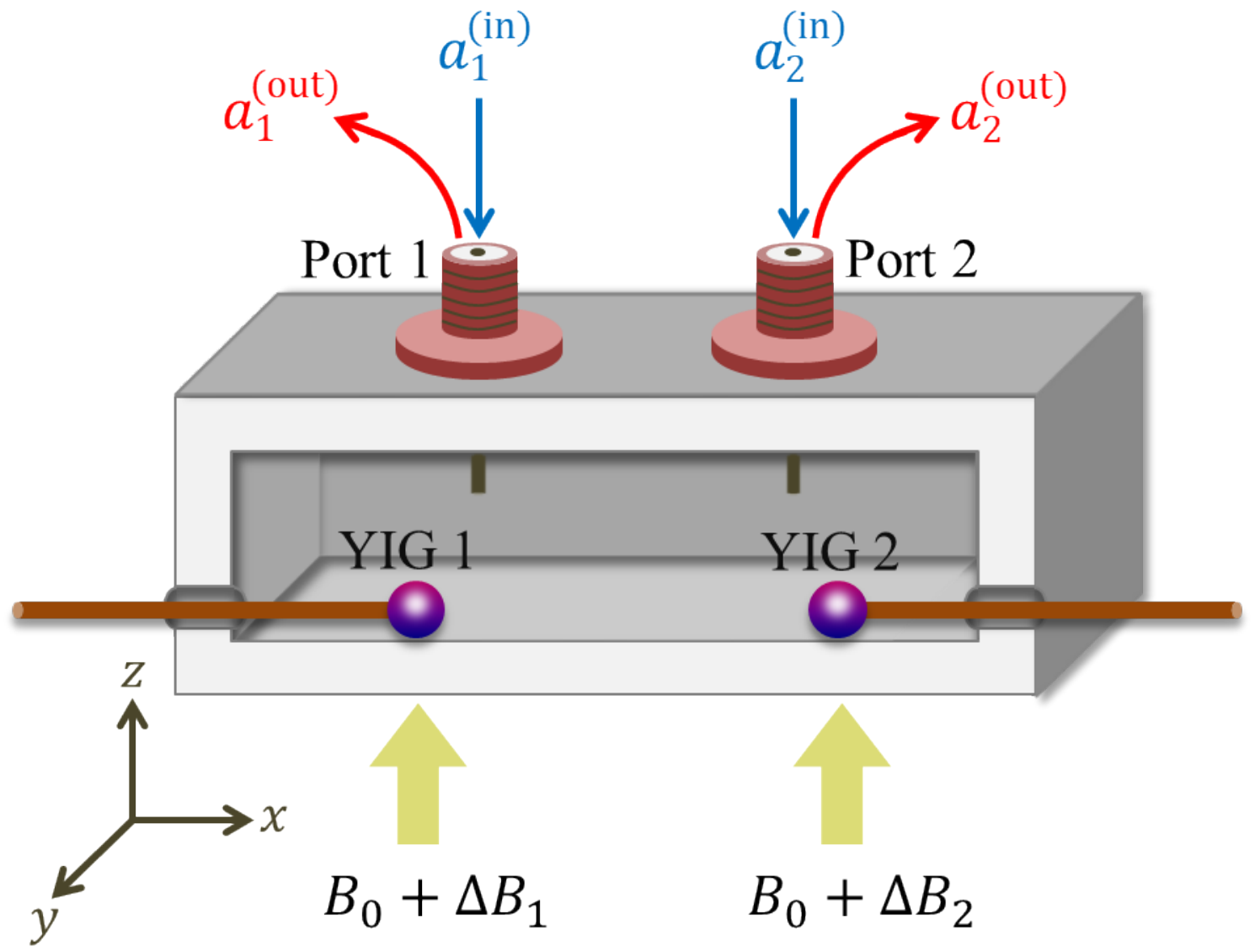}
\caption{The schematic layout of the proposed hybrid system. Two (purple) YIG spheres glued on movable (orange) thin bars are placed in a 3D microwave cavity. The magnetizations of the two YIG spheres are saturated via a static magnetic field $B_{0}$ in the $z$-direction. In addition, two weak bias magnetic fields $\Delta B_{1}$ and $\Delta B_{2}$ are also applied in the $z$-direction, each of which can be produced using a coil near the corresponding YIG sphere.~\cite{Zhang15-2} The coupling strength between each Kittel mode and the cavity mode can be controlled by moving the YIG sphere via the bar along the $x$-direction, and the decay rates $\kappa_1$ and $\kappa_2$ due to ports 1 and 2 can be tuned via changing the intracavity lengths of the pins inside the two ports.~\cite{Zhang17} Moreover, two input fields $a_{1}^{(\rm{in})}$ and $a_{2}^{(\rm{in})}$ are fed into the cavity via ports 1 and 2, and the corresponding two output fields are denoted as $a_{1}^{(\rm{out})}$ and $a_{2}^{(\rm{out})}$.}
\label{figure1}
\end{figure}

The proposed cavity magnonics system consists of two YIG spheres (YIG 1 and YIG 2) and a three-dimensional (3D) microwave cavity, as schematically shown in Fig.~\ref{figure1}, where the considered magnon mode (i.e., the Kittel mode) in each YIG sphere couples to the same cavity mode via the collective magnetic-dipole interaction. This Kittel mode corresponds to a mode of magnons in the long-wavelength limit with zero wave number (i.e., $k=0$), where all spins in the sample precess uniformly.~\cite{Kruglyak10}

When each Kittel mode is in the low-lying excitations and only one cavity mode is considered, the total Hamiltonian of this hybrid system can be written as~\cite{SoykalPRL10,SoykalPRB10,Huebl13,Tabuchi14,Zhang14}
\begin{eqnarray}\label{equ1}
\begin{split}
H=&\omega_{c}a^{\dag}a+\omega_{1}b_{1}^{\dag}b_{1}+\omega_{2}b_{2}^{\dag}b_{2}+g_{1}(a^{\dag}b_{1}+ab_{1}^{\dag})\\
   &+g_{2}(a^{\dag}b_{2}+ab_{2}^{\dag}),
\end{split}
\end{eqnarray}
where $a$ and $a^{\dag}$ ($b_{j}$ and $b_{j}^{\dag}$, $j=1, 2$) are the annihilation and creation operators of the cavity mode (the Kittel mode in the $j$th YIG sphere), $\omega_{c}$ and $\omega_{j}$ are the corresponding frequencies of these modes, and $g_{j}$ is the coupling strength between the cavity photons and the magnons in the $j$th YIG sphere. When two input fields $a_{1}^{\rm{(in)}}$ and $a_{2}^{\rm{(in)}}$ with the same frequency are fed into the microwave cavity via ports 1 and 2, the dynamics of the hybrid system is governed by the following quantum Langevin equations:~\cite{Walls94}
\begin{equation}\label{equ2}
\begin{split}
\dot{a}    =&-i\big[\omega_{c}-i(\kappa_{1}+\kappa_{2}+\kappa_{\rm{int}})\big]a -ig_{1}b_{1}-ig_{2}b_{2}\\
            &+\sqrt{2\kappa_{1}}a_{1}^{\rm{(in)}}+\sqrt{2\kappa_{2}}a_{2}^{\rm{(in)}},\\
\dot{b}_{j}=&-i(\omega_{j}-i\gamma_{j})b_{j}-ig_{j}a,
\end{split}
\end{equation}
where $\kappa_{\rm{int}}$ is the intrinsic decay rate of the cavity mode and $\kappa_{i}$ is the decay rate of the cavity mode due to the $i$th port ($i=1, 2$). Then, the total decay rate of the cavity mode is $\kappa_{1}+\kappa_{2}+\kappa_{\rm{int}}$. The Kittel mode in the $j$th YIG sphere has a damping rate $\gamma_{j}$ and no input field is applied to the Kittel mode. According to the input-output theory,~\cite{Walls94} we can connect the intra-cavity field $a$ with the input field $a_{i}^{\rm{(in)}}$ and output field $a_{i}^{\rm{(out)}}$ via
\begin{equation}\label{equ3}
a_{i}^{\rm{(in)}}+a_{i}^{\rm{(out)}}=\sqrt{2\kappa_{i}}a,
\end{equation}
at each port $i$.

\subsection{Effective Hamiltonian}

With appropriate parameters, the CPA may occur in the hybrid system (see Sec.~IIB), with no output fields going out from ports 1 and 2, i.e., $a_{i}^{\rm{(out)}}=0$. In this case, Eq.~(\ref{equ3}) becomes
\begin{equation}\label{equ4}
a_{i}^{\rm{(in)}}=\sqrt{2\kappa_{i}}a.
\end{equation}
Substituting Eq.~(\ref{equ4}) into Eq.~(\ref{equ2}), we obtain
\begin{equation}\label{equ5}
\begin{split}
\dot{a}    =&-i\big[\omega_{c}+i(\kappa_{1}+\kappa_{2}-\kappa_{\rm{int}})\big]a -ig_{1}b_{1}-ig_{2}b_{2},\\
\dot{b}_{j}=&-i(\omega_{j}-i\gamma_{j})b_{j}-ig_{j}a.
\end{split}
\end{equation}
The Langevin equations in Eq.~(\ref{equ5}) can be expressed in a matrix form as
\begin{equation}\label{equ6}
\mathbf{\dot{V}}=-iH_{\rm{eff}}\mathbf{V},
\end{equation}
where $\mathbf{V}=(a,b_{1},b_{2})^{T}$ represents a column vector and $H_{\rm{eff}}$ is the effective non-Hermitian Hamiltonian of the hybrid system,
\begin{equation}\label{equ7}
H_{\rm{eff}}=
\left(
  \begin{array}{ccc}
    \omega_{c}+i\kappa_{g} & g_{1} & g_{2}\\
    g_{1} & \omega_{1}-i\gamma_{1} & 0\\
    g_{2} & 0 & \omega_{2}-i\gamma_{2}\\
  \end{array}
\right),
\end{equation}
where $\kappa_{g}\equiv\kappa_{1}+\kappa_{2}-\kappa_{\rm{int}}> 0$ represents an effective gain of the cavity mode owing to the CPA.~\cite{Zhang17,Sun14}

In the special case without YIG 2, the effective Hamiltonian $H_{\rm{eff}}$ in Eq.~(\ref{equ7}) is reduced to a $2 \times 2$ matrix,
\begin{equation}\label{equ8}
\widetilde{H}_{\rm{eff}}=
\left(
  \begin{array}{cc}
    \omega_{c}+i\kappa_{g}    &    g_{1} \\
            g_{1}             &    \omega_{1}-i\gamma_{1} \\
  \end{array}
\right).
\end{equation}
When the system parameters satisfy $\omega_{c}=\omega_{1}$ and $\kappa_{g}=\gamma_{1}$, the binary system can posses a $\mathcal{PT}$-symmetry,~\cite{Konotop16} and the eigenvalues of $\widetilde{H}_{\rm{eff}}$ are $\omega_{\pm}=\omega_{c} \pm \sqrt{g_{1}^{2}-\gamma_{1}^{2}}$, which are real (complex) for $g_{1}>\gamma_{1}$ ($g_{1}<\gamma_{1}$), corresponding to the system in the $\mathcal{PT}$-symmetric ($\mathcal{PT}$-symmetry-breaking) phase. As reported in Ref.~\onlinecite{Zhang17}, the binary system with $\widetilde{H}_{\rm{eff}}$ exhibits a spontaneous $\mathcal{PT}$-symmetry-breaking quantum phase transition at the $\rm{EP}_{2}$ (i.e., $\omega_{+}=\omega_{-}=\omega_{c}$ when $g_{1}=\gamma_{1}$) by tuning the coupling strength $g_{1}$ from $g_{1}>\gamma_{1}$ to $g_{1}<\gamma_{1}$. In Eq.~(\ref{equ7}), when having both $g_{1} \neq 0$ and $g_{2} \neq 0$ to posses the $\mathcal{PT}$-symmetry, the ternary system should satisfy~\cite{Hodaei17} $\kappa_{g}=0$ and $\gamma_{1}=-\gamma_{2}$. This is not achievable in the usual case when the Kittel modes are lossy ($\gamma_{1} > 0$ and $\gamma_{2} > 0$). However, as shown in Sec.~III and Sec.~IV, the ternary system without the $\mathcal{PT}$-symmetry can also have the real energy spectrum and exhibit both $\rm{EP}_{3}$ and $\rm{EP}_{2}$ in the parameter space under the condition of pseudo-Hermiticity.

\subsection{CPA conditions}

Using Fourier transformations $a(t)=\frac{1}{\sqrt{2\pi}}\int^{+\infty}_{-\infty}a(\omega)e^{-i\omega t}d\omega$ and $b_{j}(t)=\frac{1}{\sqrt{2\pi}}\int^{+\infty}_{-\infty}b_{j}(\omega)e^{-i\omega t}d\omega$, we can convert the Langevin equations in Eq.~(\ref{equ2}) to
\begin{equation}\label{equ9}
\begin{split}
&-i\big[(\omega_{c}-\omega)-i(\kappa_{1}+\kappa_{2}+\kappa_{\rm{int}})\big]a -ig_{1}b_{1}-ig_{2}b_{2}\\
            &+\sqrt{2\kappa_{1}}a_{1}^{\rm{(in)}}+\sqrt{2\kappa_{2}}a_{2}^{\rm{(in)}}=0,\\
&-i\big[(\omega_{j}-\omega)-i\gamma_{j}\big]b_{j}-ig_{j}a=0.
\end{split}
\end{equation}
From Eq.~(\ref{equ9}), the intra-cavity field is obtained as
\begin{equation}\label{equ10}
a=\frac{\sqrt{2\kappa_{1}}a_{1}^{\rm{(in)}}+\sqrt{2\kappa_{2}}a_{2}^{\rm{(in)}}}
        {(\kappa_{1}+\kappa_{2}+\kappa_{\rm{int}})+i(\omega_{c}-\omega)+\sum (\omega)},
\end{equation}
where
\begin{equation}\label{equ11}
\sum (\omega)=\sum_{j=1,~2}\frac{g_{j}^{2}}{\gamma_{j}+i(\omega_{j}-\omega)}
\end{equation}
is the self-energy due to the two Kittel modes.

Using Eq.~(\ref{equ10}) and Eq.~(\ref{equ3}), we can also obtain the output fields $a_{1}^{\rm{(out)}}$ and $a_{2}^{\rm{(out)}}$ at ports 1 and 2,
\begin{equation}\label{equ12}
\begin{split}
&a_{1}^{\rm{(out)}}=\frac{2\kappa_{1}a_{1}^{\rm{(in)}}+2\sqrt{\kappa_{1}\kappa_{2}}a_{2}^{\rm{(in)}}}
        {(\kappa_{1}+\kappa_{2}+\kappa_{\rm{int}})+i(\omega_{c}-\omega)+\sum (\omega)}-a_{1}^{\rm{(in)}},\\
&a_{2}^{\rm{(out)}}=\frac{2\sqrt{\kappa_{1}\kappa_{2}}a_{1}^{\rm{(in)}}+2\kappa_{2}a_{2}^{\rm{(in)}}}
        {(\kappa_{1}+\kappa_{2}+\kappa_{\rm{int}})+i(\omega_{c}-\omega)+\sum (\omega)}-a_{2}^{\rm{(in)}}.
\end{split}
\end{equation}
When the CPA occurs, the two input fields are fully fed into the cavity, so $a_{1}^{\rm{(out)}}=a_{2}^{\rm{(out)}}=0$ in Eq.~(\ref{equ12}). Solving Eq.~(\ref{equ12}) with $a_{i}^{\rm{(out)}}=0$, we obtain three constraints:

The first constraint on the two input fields $a_{1}^{\rm{(in)}}$ and $a_{2}^{\rm{(in)}}$ is
\begin{equation}\label{equ13}
a_{2}^{\rm{(in)}} = \sqrt{\kappa_{2}/\kappa_{1}}a_{1}^{\rm{(in)}},
\end{equation}
while the second and third constraints on the parameters of the system and the frequency of the input fields are
\begin{eqnarray}\label{equ14}
\begin{split}
\kappa_{g}                   = & \sum_{j=1,~2}\frac{g_{j}^{2}}{(\omega_{j}-\omega_{\rm{CPA}})^{2}+\gamma_{j}^{2}}\gamma_{j},\\
\omega_{c}-\omega_{\rm{CPA}} = & \sum_{j=1,~2}\frac{g_{j}^{2}}{(\omega_{j}-\omega_{\rm{CPA}})^{2}+\gamma_{j}^{2}}(\omega_{j}-\omega_{\rm{CPA}}),
\end{split}
\end{eqnarray}
where $\omega_{\rm{CPA}}$ denotes the frequency of the two input fields in the case of the CPA.  The constraint in Eq.~(\ref{equ13}) means that the two input fields should have the same phase and a specific magnitude ratio $\sqrt{\kappa_{2}/\kappa_{1}}$, which can be readily satisfied via a variable phase shifter and a variable attenuator in the experiment.~\cite{Zhang17}

\section{Pseudo-Hermitian Hamiltonian}\label{general}

 Below we derive the parameter conditions of the pseudo-Hermiticity for the effective Hamiltonian $H_{\rm{eff}}$ in Eq.~(\ref{equ7}). For this considered Hamiltonian, there are three eigenvalues. Following Ref.~\onlinecite{Mostafazadeh02-1}, $H_{\rm{eff}}$ becomes a pseudo-Hermitian only if its eigenvalues satisfy one of the following conditions: (i) all three eigenvalues are real, or (ii) one of the three eigenvalues is real and other two are a complex-conjugate pair. Solving ${\rm Det}(H_{\rm{eff}}-\Omega I)=0$, i.e.,
\begin{equation}\label{equ15}
\begin{split}
\left|
  \begin{array}{ccc}
    (\omega_{c}+i\kappa_{g})-\Omega & g_{1} & g_{2}\\
    g_{1} & (\omega_{1}-i\gamma_{1})-\Omega & 0\\
    g_{2} & 0 & (\omega_{2}-i\gamma_{2})-\Omega\\
  \end{array}
\right|=0,
\end{split}
\end{equation}
where $I$ is an identity matrix, we can obtain the three eigenvalues.
According to the energy-spectrum property of the pseudo-Hermitian Hamiltonian,~\cite{Mostafazadeh02-1} both Eq.~(\ref{equ15}) and its complex-conjugate expression ${\rm Det}(H_{\rm{eff}}^{*}-\Omega I)=0$, i.e.,
\begin{equation}\label{equ16}
\begin{split}
\left|
  \begin{array}{ccc}
    (\omega_{c}-i\kappa_{g})-\Omega & g_{1} & g_{2}\\
    g_{1} & (\omega_{1}+i\gamma_{1})-\Omega & 0\\
    g_{2} & 0 & (\omega_{2}+i\gamma_{2})-\Omega\\
  \end{array}
\right|=0,
\end{split}
\end{equation}
should yield the same solutions.

By expanding the determinants in Eqs.~(\ref{equ15}) and (\ref{equ16}) and comparing their corresponding coefficients, we find that the system parameters satisfy the following constraints:
\begin{equation}\label{equ17}
\begin{split}
&\kappa_{g}-\gamma_{1}-\gamma_{2}=0,\\
&\Delta_{1}\gamma_{1}+\Delta_{2}\gamma_{2}=0,\\
&(\Delta_{1}\Delta_{2}-\gamma_{1}\gamma_{2})\kappa_{g}+g_{1}^{2}\gamma_{2}+g_{2}^{2}\gamma_{1}=0,
\end{split}
\end{equation}
and the characteristic polynomial in Eq.~(\ref{equ15}) is reduced to
\begin{equation}\label{equ18}
(\Omega-\omega_{c})^{3}+c_{2}(\Omega-\omega_{c})^{2}+c_{1}(\Omega-\omega_{c})+c_{0}=0.
\end{equation}
Here $\Delta_{1(2)}=\omega_{1(2)}-\omega_{c}$ is the frequency detuning between the Kittel mode 1 (2) and the cavity mode, and the coefficients $c_{0}$, $c_{1}$ and $c_{2}$ are given by
\begin{equation}\label{equ19}
\begin{split}
&c_{0}=g_{1}^{2}\Delta_{2}+g_{2}^{2}\Delta_{1}-\kappa_{g}(\gamma_{1}\Delta_{2}+\gamma_{2}\Delta_{1}),\\
&c_{1}=\kappa_{g}^{2}+\Delta_{1}\Delta_{2}-\gamma_{1}\gamma_{2}-g_{1}^{2}-g_{2}^{2},\\
&c_{2}=-(\Delta_{1}+\Delta_{2}).
\end{split}
\end{equation}
Clearly, the pseudo-Hermiticity ensures that the loss and gain are balanced in the whole hybrid system, i.e., $\kappa_{g}-\gamma_{1}-\gamma_{2}=0$. For convenience, we introduce two new parameters $\eta$ and $k$,
\begin{equation}\label{equ20}
\gamma_{1}=\eta\gamma_{2},~~~g_{2}=kg_{1},
\end{equation}
where we have assumed that $\gamma_{2} \leq \gamma_{1}$, i.e., $\eta \geq 1$. Using Eq.~(\ref{equ20}), the pseudo-Hermitian conditions in Eq.~(\ref{equ17}) become
\begin{eqnarray}\label{equ21}
\kappa_{g}&=&(1+\eta)\gamma_{2},\nonumber\\
\Delta_{2}&=&-\eta \Delta_{1},\\
\Delta_{1}^{2}&=&\frac{1+\eta k^{2}}{(1+\eta)\eta}g_{1}^{2}-\gamma_{2}^{2},\nonumber
\end{eqnarray}
and the coefficients of the characteristic polynomial in Eq.~(\ref{equ19}) are
\begin{equation}\label{equ22}
\begin{split}
&c_{0}=(k^{2}-\eta)g_{1}^{2}\Delta_{1}+(\eta^{2}-1)(1+\eta)\gamma_{2}^{2}\Delta_{1},\\
&c_{1}=(1+\eta)^{2}\gamma_{2}^{2}-\eta(\Delta_{1}^{2}+\gamma_{2}^{2})-(1+k^{2})g_{1}^{2},\\
&c_{2}=(\eta-1)\Delta_{1}.
\end{split}
\end{equation}
From the last equation in Eq.~(\ref{equ21}), it follows that the coupling strength $g_{1}$ should be in an appropriate regime to ensure $\Delta_{1}^{2} \geq 0$. Setting $\Delta_{1}^{2}$=0, the allowed minimal value $g_{\rm{min}}$ of the coupling strength $g_{1}$ is given by
\begin{equation}\label{equ23}
g_{\rm{min}} \equiv \bigg[\frac{(1+\eta)\eta}{1+\eta k^{2}}\bigg]^{1/2} \gamma_{2}.
\end{equation}
Obviously, this is achievable in our considered system.

\section{$\text{EP}_{3}$ in the cavity magnonics system}\label{EP3}

In this section, we study the $\text{EP}_{3}$ in both symmetric and asymmetric cases by solving the characteristic polynomial in Eq.~(\ref{equ18}) under the pseudo-Hermitian conditions of the system's parameters and demonstrate that this $\text{EP}_{3}$ can be observable via measuring the total output spectrum of the cavity. Assuming that the pseudo-Hermitian system has an $\text{EP}_{3}$ at $\Omega \equiv \Omega_{\rm{EP3}}$ and the corresponding critical parameters are denoted as $g_{1} \equiv g_{\rm{EP3}}$ and $\Delta_{1} \equiv \Delta_{\rm{EP3}}$, we can rewrite the secular equation in Eq.~(\ref{equ18}) as
\begin{equation}\label{equ24}
(\Omega-\Omega_{\rm{EP3}})^{3}=0
\end{equation}
at the $\text{EP}_{3}$. Comparing the coefficients of Eq.~(\ref{equ18}) and Eq.~(\ref{equ24}), we can link the coalescence eigenvalue $\Omega=\Omega_{\rm{EP3}}$ to the parameters of the system,
\begin{equation}\label{equ25}
\begin{split}
-3(\Omega_{\rm{EP3}}-\omega_{c})   =\,&(\eta-1)\Delta_{\rm{EP3}},\\
3(\Omega_{\rm{EP3}}-\omega_{c})^{2}=\,&(1+\eta)^{2}\gamma_{2}^{2}-\eta(\Delta_{\rm{EP3}}^{2}+\gamma_{2}^{2})-(1+k^{2})g_{\rm{EP3}}^{2},\\
-(\Omega_{\rm{EP3}}-\omega_{c})^{3}=\,&(k^{2}-\eta)g_{\rm{EP3}}^{2}\Delta_{\rm{EP3}}+(\eta^{2}-1)(1+\eta)\gamma_{2}^{2}\Delta_{\rm{EP3}}.
\end{split}
\end{equation}
The first equation in Eq.~(\ref{equ25}) gives the corresponding eigenvalue at the $\text{EP}_{3}$,
\begin{equation}\label{equ26}
\Omega_{\rm{EP3}}=\omega_{c}+\frac{1}{3}(1-\eta)\Delta_{\rm{EP3}}.
\end{equation}

\subsection{The symmetric case of $\gamma_{1}=\gamma_{2}$}

When the two Kittel modes have identical damping rates $\gamma_{1}=\gamma_{2}$ (i.e., $\eta=1$), the coalescence eigenvalue in Eq.~(\ref{equ26}) becomes $\Omega_{\rm{EP3}}=\omega_{c}$ and the last two equations in Eq.~(\ref{equ25}) can be simplified to
\begin{equation}\label{equ27}
\begin{split}
&\Delta_{\rm{EP3}}^{2}+(1+k^{2})g_{\rm{EP3}}^{2}-3\gamma_{2}^{2}=0,\\
&(k^{2}-1)g_{\rm{EP3}}^{2}\Delta_{\rm{EP3}}=0.
\end{split}
\end{equation}
Solving Eq.~(\ref{equ27}) under the pseudo-Hermitian conditions in Eq.~(\ref{equ21}) and ignoring the trivial solution, we can analytically express the critical parameters as
\begin{equation}\label{equ28}
g_{\rm{EP3}}=\frac{2}{\sqrt{3}}\gamma_{2},~~~ \Delta_{\rm{EP3}}=\frac{1}{\sqrt{3}}\gamma_{2},
\end{equation}
and the obtained ratio $k$ in Eq.~(\ref{equ20}) is $k=1$.

In such a case with $\gamma_{1}=\gamma_{2}$ and $g_{1}=g_{2}$ (i.e., $\eta=k=1$), the secular equation in Eq.~(\ref{equ18}) can be rewritten as
\begin{equation}\label{equ29}
\big[(\Omega-\omega_{c})^{2}-(3g_{1}^{2}-4\gamma_{2}^{2})\big](\Omega-\omega_{c})=0.
\end{equation}
The corresponding three eigenvalues of the effective pseudo-Hermitian Hamiltonian $H_{\rm{eff}}$ are
\begin{equation}\label{equ30}
\begin{split}
&\Omega_{0}=\omega_{c},\\
&\Omega_{\pm}=\omega_{c} \pm \sqrt{3g_{1}^{2}-4\gamma_{2}^{2}},
\end{split}
\end{equation}
in the region $g_{1} \geq g_{\rm{min}}$. Now, $g_{\rm{min}}$ in Eq.~(\ref{equ23}) becomes $g_{\rm{min}}=\gamma_2$, which is smaller than $g_{\rm{EP3}}=\frac{2}{\sqrt{3}}\gamma_{2}$.
Clearly, the eigenvalue $\Omega_{0}$ is real for any allowed values of $g_{1}$ (i.e., $g_{1} \geq g_{\rm{min}}$), while the two eigenvalues $\Omega_{\pm}$ are real for $3g_{1}^{2}-4\gamma_{2}^{2} > 0$ (i.e., $g_{1} > g_{\rm{EP3}}$) and complex for $3g_{1}^{2}-4\gamma_{2}^{2} < 0$ (i.e., $g_{\rm{min}}\leq g_{1} < g_{\rm{EP3}}$). When $3g_{1}^{2}-4\gamma_{2}^{2}=0$ (i.e., $g_{1}=g_{\rm{EP3}}$), the three eigenvalues $\Omega_{\pm}$ and $\Omega_{0}$ coalesce to the $\rm{EP}_{3}$, i.e., $\Omega_{\pm}=\Omega_{0}=\Omega_{\rm{EP3}}=\omega_c$. Because $g_{\rm{min}}<g_{\rm{EP3}}$, the $\text{EP}_{3}$ is experimentally observable in this symmetric case.

Below we check the CPA conditions in Eq.~(\ref{equ14}). For $\eta=k=1$, the CPA conditions are reduced to
\begin{equation}\label{equ31}
\begin{split}
&\big[(\omega_{\rm{CPA}}-\omega_{c})^{2}-(3g_{1}^{2}-4\gamma_{2}^{2})\big](\omega_{\rm{CPA}}-\omega_{c})^{2}=0,\\
&\big[(\omega_{\rm{CPA}}-\omega_{c})^{2}-(3g_{1}^{2}-4\gamma_{2}^{2})\big]\big[(\omega_{\rm{CPA}}-\omega_{c})^{2}-g_{1}^{2}\big]\\
& \times     (\omega_{\rm{CPA}}-\omega_{c})=0,
\end{split}
\end{equation}
under the pseudo-Hermitian conditions in Eq.~(\ref{equ21}). Solving the above equations, we obtain the three CPA frequencies
\begin{equation}\label{equ32}
\begin{split}
&\omega_{\rm{CPA}}^{(0)}=\omega_{c},~~~~~~g_{1} \geq g_{\rm{min}};\\
&\omega_{\rm{CPA}}^{(\pm)}=\omega_{c} \pm \sqrt{3g_{1}^{2}-4\gamma_{2}^{2}},~~~~~~g_{1} \geq g_{\rm{EP3}}.
\end{split}
\end{equation}
Comparing Eq.~(\ref{equ32}) with Eq.~(\ref{equ30}), we find that the CPA frequencies are coincident with the eigenvalues of the hybrid system when the eigenvalues $\Omega_{\pm}$ and $\Omega_{0}$ are real. However, for the complex eigenvalues, the CPA goes to disappear.

\begin{figure}
\includegraphics[width=0.4\textwidth]{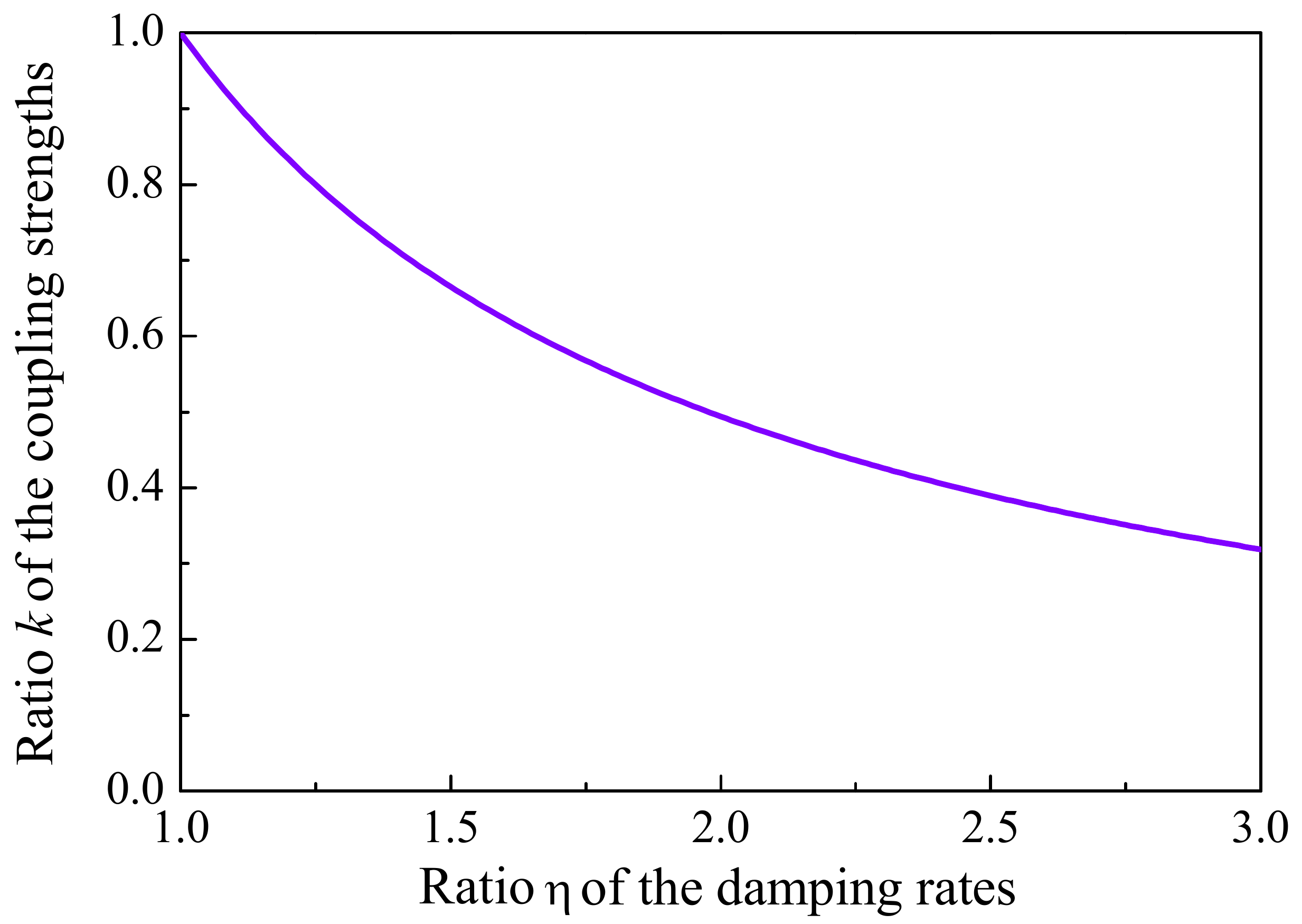}
\caption{The ratio $k=g_{2}/g_{1}$ of the coupling strengths $g_{2}$ and $g_{1}$ versus the ratio $\eta=\gamma_{1}/\gamma_{2}$ of the Kittel-mode damping rates $\gamma_{1}$ and $\gamma_{2}$.}
\label{figure2}
\end{figure}

\subsection{The asymmetric case of $\gamma_{1} \neq \gamma_{2}$}

\begin{figure}
\includegraphics[width=0.48\textwidth]{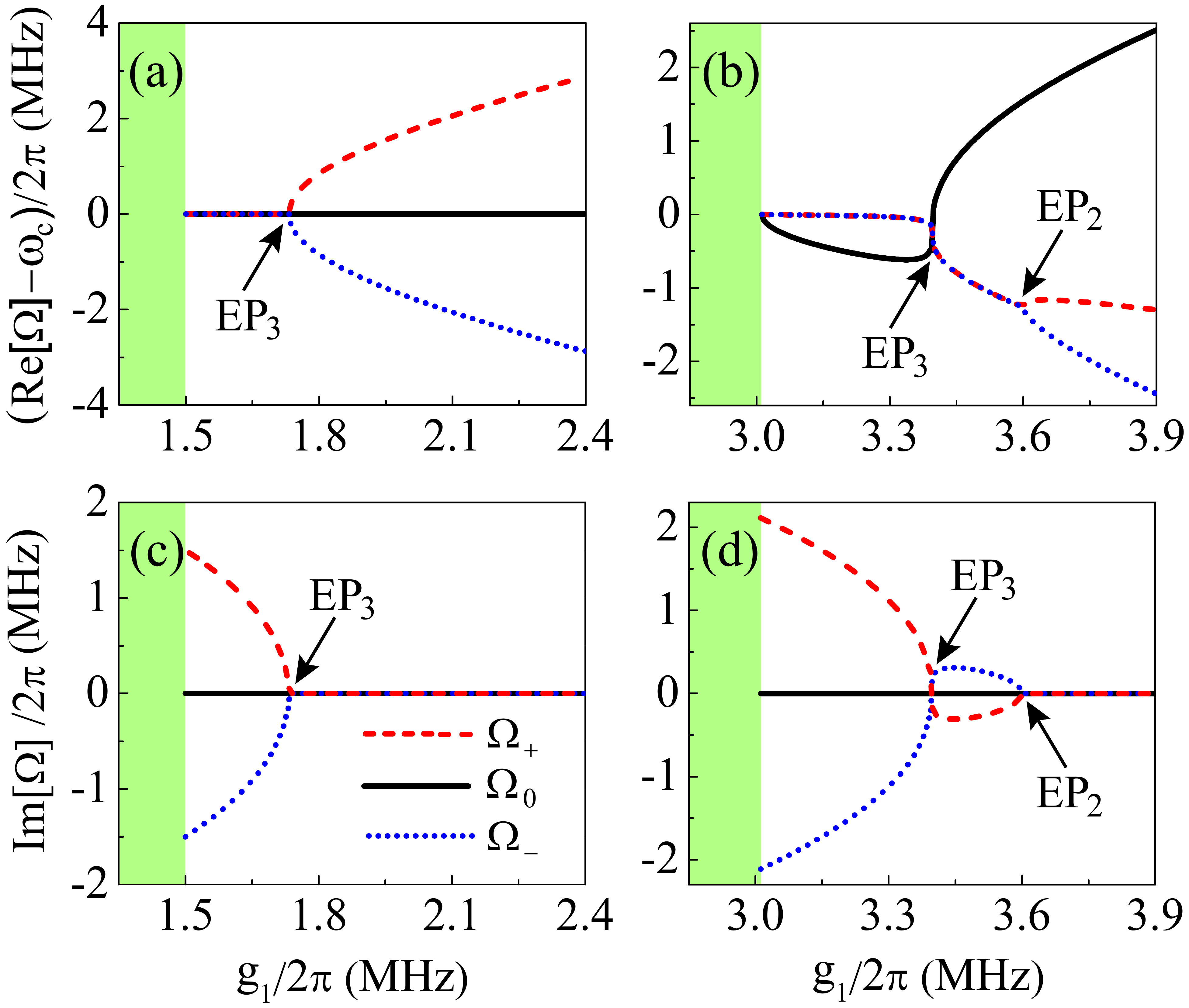}
\caption{The eigenvalues of the effective Hamiltonian $H_{\rm{eff}}$ in Eq.~(\ref{equ7}) versus the coupling strength $g_{1}$ between the cavity mode and the Kittel mode in YIG 1. Note that there is no pseudo-Hermiticity for the system with $g_{1} < g_{\rm{min}}$ (green regions). In each figure, the dashed red and dotted blue lines denote the eigenvalues $\Omega_{\pm}$ and the solid black line denotes the eigenvalue $\Omega_{0}$. (a) and (c) The real and imaginary parts of the eigenvalues $\Omega_{\pm}$ and $\Omega_{0}$ versus $g_{1}$ in the symmetric case of $\eta=k=1$, where $\kappa_{1}/2\pi=\kappa_{2}/2\pi=2.25$ MHz and $\gamma_{1}/2\pi=1.5$ MHz. (b) and (d) The real and imaginary parts of the eigenvalues $\Omega_{\pm}$ and $\Omega_{0}$ versus $g_{1}$ in the asymmetric case of $\eta=2$ and $k=0.494$, where $\kappa_{1}/2\pi=\kappa_{2}/2\pi=3$ MHz and $\gamma_{1}/2\pi=3$ MHz. Other parameters are chosen to be $\gamma_{2}/2\pi=\kappa_{\rm{int}}/2\pi=1.5$ MHz.}
\label{figure3}
\end{figure}

In the experiment, it is difficult to have two Kittel modes with the same damping rates, because the Kittel-mode damping rate is not tunable. Thus, it is useful to investigate the $\text{EP}_{3}$ in the asymmetric case of $\gamma_{1} \neq \gamma_{2}$ (i.e., $\eta \neq 1$). With the pseudo-Hermitian conditions in Eq.~(\ref{equ21}) and the conditions of the $\text{EP}_{3}$ in Eq.~(\ref{equ25}), we find that the parameter $k$ satisfies the following expression:
\begin{equation}\label{equ33}
\begin{split}
&\frac{1}{4}\bigg[\frac{1+\eta k^{2}}{(1+\eta)\eta}+\frac{3(1+k^{2})}{1+\eta+\eta^{2}}\bigg]\\
&=\bigg[1+\frac{27(1+\eta)^{2}}{(\eta-1)^{2}}\bigg]^{-1}
           \bigg[\frac{1+\eta k^{2}}{(1+\eta)\eta}-\frac{27(k^{2}-\eta)}{(\eta-1)^{3}}\bigg],
\end{split}
\end{equation}
where $\eta\neq 1$, and the critical parameters are
\begin{equation}\label{equ34}
\begin{split}
&g_{\rm{EP3}}=\bigg[\frac{1+\eta k^{2}}{(1+\eta)\eta}+\frac{3(1+k^{2})}{1+\eta+\eta^{2}}\bigg]^{-1/2}2\gamma_{2},\\
&\Delta_{\rm{EP3}}=\bigg[\frac{1+\eta k^{2}}{(1+\eta)\eta}g_{\rm{EP3}}^{2}-\gamma_{2}^{2}\bigg]^{1/2}.
\end{split}
\end{equation}
With the expressions of $g_{\rm{min}}$ and $g_{\rm{EP3}}$ in Eqs.~(\ref{equ23}) and (\ref{equ34}), respectively, we obtain the relation
\begin{equation}
g_{\rm{EP3}}=g_{\rm{min}}\sqrt{1+27\eta^{2}(2+\eta)^{-2}(1+2\eta)^{-2}},
\end{equation}
where we eliminate the parameter $k$ via Eq.~(\ref{equ33}). Obviously, $g_{\rm{EP3}}>g_{\rm{min}}$, so the $\rm{EP}_{3}$ is achievable in the experiment.

Using Eq.~(\ref{equ33}), we plot in Fig.~\ref{figure2} the ratio $k=g_{2}/g_{1}$ of the coupling strengths versus the ratio $\eta=\gamma_{1}/\gamma_{2}$ of the damping rates. It can be seen that $k$ decreases from 1 to 0.3 as $\eta$ varies from 1 to 3, which means that the coupling strengths should satisfy the relation $g_{1} > g_{2}$ in the case of $\gamma_{1} > \gamma_{2}$ (because $\gamma_{1}=\eta\gamma_{2}$ and $g_{1}=g_{2}/k$) to observe $\rm{EP}_{3}$ in our proposed system. Different from the symmetric case with $\gamma_{1}=\gamma_{2}$ and $g_{1}=g_{2}$ (i.e., $\eta=k=1$), it is difficult to analytically solve the secular equation in Eq.~(\ref{equ18}) as well as the CPA conditions in Eq.~(\ref{equ14}) for $\gamma_{1} \neq \gamma_{2}$ and $g_{1} \neq g_{2}$, but we can numerically solve them.

In Fig.~\ref{figure3}, we plot the energy spectra of the effective Hamiltonian $H_{\rm{eff}}$ in Eq.~(\ref{equ7}) versus the coupling strength $g_{1}$ in the symmetric and asymmetric cases of $\gamma_{1} =\gamma_{2}$ and $\gamma_{1}=2\gamma_{2}$ (i.e., $\eta=1$ and $\eta=2$), respectively. Note that no eigenvalue exists when $g_{1} < g_{\rm{min}}$ (see the green regions), because there is no pseudo-Hermiticity for the system. Figures~\ref{figure3}(a) and \ref{figure3}(c) show the real and imaginary parts of the eigenvalues $\Omega_{\pm}$ and $\Omega_{0}$ given in Eq.~(\ref{equ30}) versus $g_{1}$ for $\eta=1$, with the critical coupling strength $g_{\rm{EP3}}/2\pi = 1.732$ MHz. The eigenvalues have different characteristics in the two regions: $g_{\rm{min}}\leq g_{1} < g_{\rm{EP3}}$ and $g_{1} > g_{\rm{EP3}}$. When $g_{\rm{min}}\leq g_{1} < g_{\rm{EP3}}$, the eigenvalues $\Omega_{\pm}$ are a complex-conjugate pair (see the dashed red and dotted blue lines) and $\Omega_{0}$ is real (see the solid black lines). It is clear that the three eigenvalues $\Omega_{\pm}$ and $\Omega_{0}$ coalesce to $\Omega_{\rm{EP3}}=\omega_{c}$ at $g_{1}=g_{\rm{EP3}}$ (i.e., the $\text{EP}_{3}$). For $g_{1} > g_{\rm{EP3}}$, all three eigenvalues $\Omega_{\pm}$ and $\Omega_{0}$ are real.

In the asymmetric case of $\gamma_{1} \neq \gamma_{2}$ (where we choose $\eta=2$), the corresponding real and imaginary parts of the eigenvalues $\Omega_{\pm}$ and $\Omega_{0}$ versus the coupling strength $g_{1}$ are shown in Figs.~\ref{figure3}(b) and \ref{figure3}(d), respectively. We note that there are two critical coupling strengths $g_{\rm{EP3}}/2\pi = 3.394$ MHz and $g_{\rm{EP2}}/2\pi = 3.600$ MHz. The eigenvalue $\Omega_{0}$ is real for any allowed values of the coupling strength $g_{1}\geq g_{\rm{min}}$ (see the solid black lines) and $\Omega_{\pm}$ are complex (real) for $g_{\rm{min}}\leq g_{1} < g_{\rm{EP3}}$ and $g_{\rm{EP3}} < g_{1} < g_{\rm{EP2}}$ ($g_{1}=g_{\rm{EP3}}$ and $g_{1} \geq g_{\rm{EP2}}$) (see the dashed red and dotted blue lines). In this case, in addition to the $\rm{EP}_{3}$ at $g_{1} = g_{\rm{EP3}}$, where the three eigenvalues $\Omega_{\pm}$ and $\Omega_{0}$ are coalescent, there is the $\rm{EP}_{2}$ at $g_{1} = g_{\rm{EP2}}$, where the two eigenvalues $\Omega_{\pm}$ are coalescent. This is different from the symmetric case.

\subsection{The output spectrum}

In this subsection, we derive the total output spectrum of the cavity for the hybrid system and show that the pseudo-Hermiticity can be observed using the output spectrum. As discussed in Sec.~IIB, when the CPA occurs, the first constraint is on the two input fields $a_{1}^{\rm{(in)}}$ and $a_{2}^{\rm{(in)}}$, i.e., Eq.~(\ref{equ13}). Using this equation, the expressions of the two outgoing fields in Eq.~(\ref{equ12}) can be rewritten as
\begin{equation}\label{equ35}
\begin{split}
&a_{1}^{\rm{(out)}}=S_{1}(\omega)a_{1}^{\rm{(in)}},\\
&a_{2}^{\rm{(out)}}=S_{2}(\omega)a_{2}^{\rm{(in)}},
\end{split}
\end{equation}
where $S_{1}(\omega)$ and $S_{2}(\omega)$ are the output coefficients at ports 1 and 2 for the frequency $\omega$ of the two input fields,
\begin{equation}\label{equ36}
\begin{split}
&S_{1}(\omega)=
\frac{2\kappa_{1}+2\kappa_{2}}{(\kappa_{1}+\kappa_{2}+\kappa_{\rm{int}})+i(\omega_{c}-\omega)+\sum (\omega)}-1,\\
&S_{2}(\omega)=S_{1}(\omega).
\end{split}
\end{equation}
Here we define a total output spectrum $|S_{\rm{tot}}(\omega)|^2$ to characterize the input-output property of the hybrid system,
\begin{equation}\label{equ37}
|S_{\rm{tot}}(\omega)|^{2}=|S_{1}(\omega)|^{2}+|S_{2}(\omega)|^{2}.
\end{equation}
It is easy to check that $|S_{\rm{tot}}(\omega)|^2=0$ when the second and third constraints in Eq.~(\ref{equ14}) are satisfied at $\omega=\omega_{\rm CPA}$.

\begin{figure}
\includegraphics[width=0.45\textwidth]{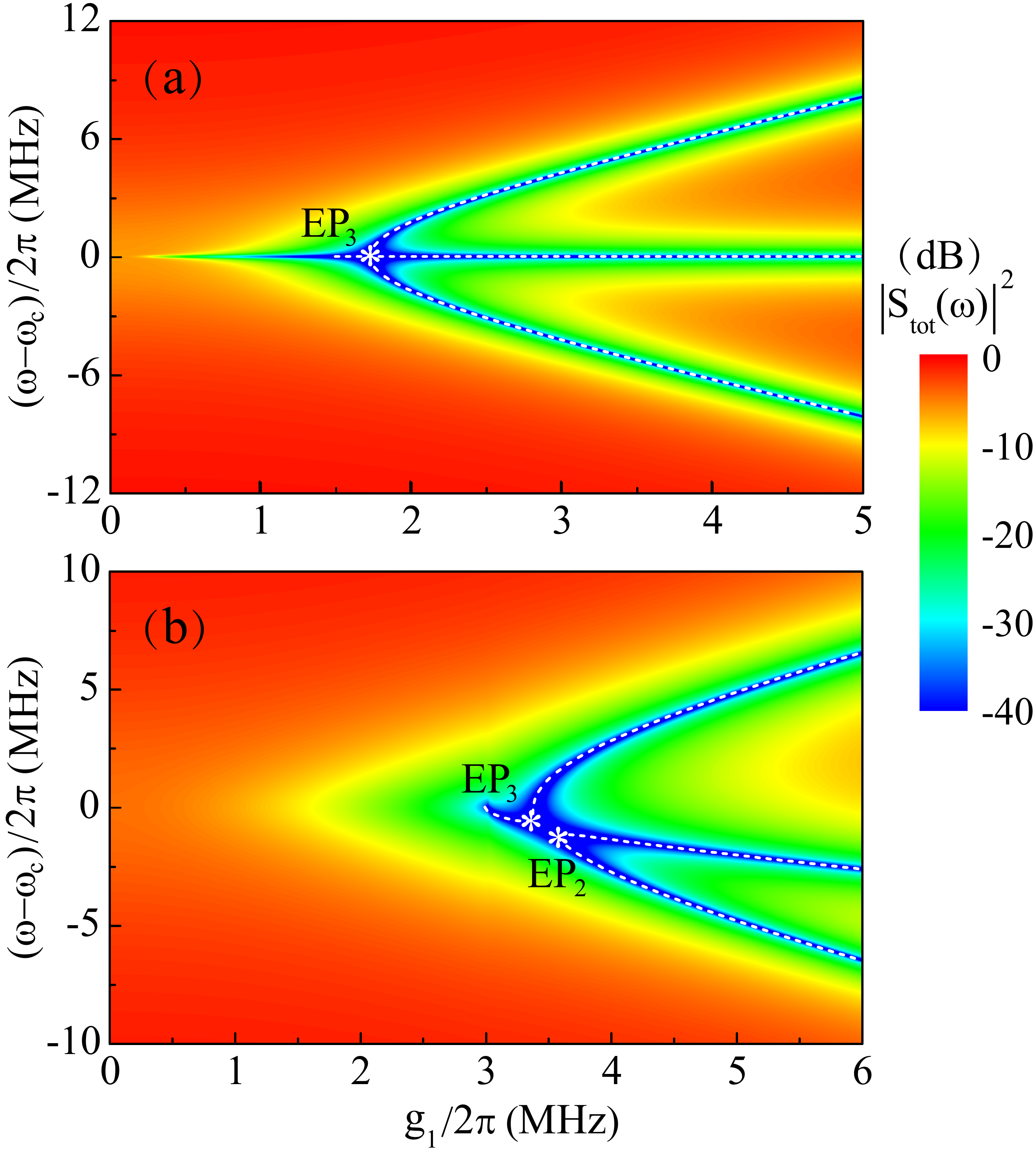}
\caption{The transmission spectrum $|S_{\rm{tot}}(\omega)|^{2}$ versus the coupling strength $g_{1}$ between the cavity mode and the Kittel mode in YIG 1 as well as the frequency detuning $\omega-\omega_{c}$ between the input fields and the cavity mode, where the phases and amplitudes of the two input fields satisfy Eq.~(\ref{equ13}). (a) The symmetric case of $\gamma_{1}=\gamma_{2}$ (i.e., $\eta=1$). (b) The asymmetric case of $\gamma_{1}=2\gamma_{2}$ (i.e., $\eta=2$). The real energy spectra (cf. Fig.~\ref{figure3}) and EPs are marked using the dashed white lines and the white stars, respectively. Other parameters are the same as in Fig.~\ref{figure3}.}
\label{figure4}
\end{figure}

In Figs.~\ref{figure4}(a) and \ref{figure4}(b), we show the total output spectrum $|S_{\rm{tot}}(\omega)|^{2}$ versus the coupling strength $g_{1}$ and the frequency detuning $\omega-\omega_{c}$ between the two input fields and the cavity mode, when $\eta=1$ and $\eta=2$, respectively. The minimum in the total output spectrum (see the blue pattern) represents the CPA, i.e., $a_{1}^{\rm{(out)}}=a_{2}^{\rm{(out)}}=0$. As expected, the CPA frequencies are coincident with the real eigenfrequencies of the effective pseudo-Hermitian Hamiltonian $H_{\rm{eff}}$ in Eq.~(\ref{equ7}), where the real eigenvalues and the EPs are indicated by the dashed white lines and the white stars, respectively. Therefore, the energy spectra as well as the $\rm{EP}_{3}$ and $\rm{EP}_{2}$ can be demonstrated via measuring the total output spectrum of the microwave cavity.

\section{Discussions and conclusions}\label{discussion}

Both the 3D microwave cavity with a high $Q$ factor (e.g., $\kappa_{\rm{int}}/2\pi \sim $ 1 MHz) and the highly-polished small YIG sphere with $\gamma_{1,2}/2\pi \sim $ 1 MHz are experimentally available.~\cite{Tabuchi14,Zhang14,Zhang15-1} Also, the decay rates of the cavity induced by the two ports are tunable (ranging from 0 to 8 MHz) by adjusting the intracavity pin lengths of the ports.~\cite{Zhang17} For a saturated magnetized YIG sphere by a static magnetic field, the frequency of the Kittel mode in the YIG sphere can be further tuned in the range of tens of megahertz via the magnetic field generated by a small coil near the sphere.~\cite{Zhang15-2} In Ref.~\onlinecite{Zhang17}, the YIG sphere is adhered to a thin rod placed into the cavity through a small hole of the cavity and the coupling strength between the cavity and Kittel mode can be tuned from $0$ to $9$ MHz by moving the rod. Moreover, a microwave signal generated by a vector network analyzer can be divided into two feeding fields needed for realizing the CPA and their magnitudes and relative phases can be adjusted using variable attenuator and phase shifter, respectively. With these achievable conditions, our proposed scheme is experimentally implementable.

In addition, we have shown that the energy spectra and $\rm{EP}_{3}$ can be revealed by harnessing the output spectrum of the microwave cavity. Owing to the good tunability of the cavity magnonics system, the CPA conditions can be nearly perfectly satisfied by carefully adjusting the parameters of the system,~\cite{Zhang17} where the absorption rate of the cavity for the two input fields can reach as high as $99\%$. Indeed, as shown in a cavity magnonics system with only one YIG sphere,~\cite{Zhang17} the experimentally obtained energy spectra and EP can have the similar features as the simulated counterparts of the ideal system, even though the CPA conditions cannot be ideally achieved in the experiment. This confirms the usefulness of the spectroscopic method in revealing the exceptional points.

In Ref.~\onlinecite{Zhang17}, the CPA was achieved for a $\mathcal{PT}$-symmetric system, while in Ref.~\onlinecite{Mostafazadeh12} the CPA was investigated for an optical system without the $\mathcal{PT}$ symmetry. In the present work, we find that the CPA is also realizable for a system with the pseudo-Hermiticity. It is known that the $\rm{EP}_{3}$ is more complicated but has richer physics than the $\rm{EP}_{2}$.~\cite{Lin16,Graefe08,Ryu12,Heiss15,Schnabel17,Jing17,Wu18,Ding16,Hodaei17} Compared with other platforms, the cavity magnonics system has its own merits, such as the high tunability and good coherence,~\cite{Zhang15-2,Wang16,Wang18,Zhang17} which are important for the implementation of the pseudo-Hermiticity of the system. Also, as a hybrid system, the cavity magnonics system has good compatibility with phonons,~\cite{Zhang16-1} optical photons~\cite{Haigh15,Osada16,Zhang16-2,Haigh16} and superconducting qubits.~\cite{Tabuchi15,Quirion17} Moreover, the YIG has the intrinsic nonreciprocity.~\cite{Osada16} These characteristics will make the cavity magnonics system useful in exploring the richer properties of the high-order exceptional points.

In short, we have theoretically studied the pseudo-Hermiticity and $\rm{EP}_{3}$ in a cavity magnonics system consisting of two small YIG spheres in a microwave cavity. Under the parameter conditions of the pseudo-Hermiticity, the effective Hamiltonian of the system has either three real eigenvalues or one real and two complex-conjugate eigenvalues. By tuning the coupling strengths between the two Kittel modes and the cavity mode, the three eigenvalues can coalesce at the $\rm{EP}_{3}$. Also, we show that the pseudo-Hermiticity and $\rm{EP}_{3}$ can be probed using the total output spectrum of the cavity.
Our work provides an experimentally feasible scheme to realize the pseudo-Hermiticity and $\rm{EP}_{3}$ in a hybrid quantum system.

\section*{Acknowledgments}

We thank P. Rabl for useful discussions. This work is supported by the National Key Research and Development Program of China (grant No.~2016YFA0301200) and the National Natural Science Foundation of China (NSFC) (Grant Nos.~U1801661, 11774022, and U1530401).

\end{document}